\documentclass[
amssymb,%
12pt,%
final,%
notitlepage,%
oneside,%
onecolumn,%
nobibnotes,%
nofootinbib,%
superscriptaddress,%
noshowpacs,%
centertags]%
{revtex4}
\usepackage{graphicx}

\begin{document}

\title{Total Cross Section, Inelasticity and Multiplicity Distributions in Proton -- Proton Collisions.}
\author{\firstname{G.} \surname{Musulmanbekov}}

%
\begin{abstract}
Multiparticle production in high energy proton -- proton collisions has
been analysed in the frame of Strongly Correlated Quark Model (SCQM) of
the hadron structure elaborated by the author. It is shown that
inelasticity decreases at high energies and this effect together with the
total cross section growth and the increasing with collision energy the
masses of intermediate clusters result in the violation of KNO -- scaling.
\end{abstract}
\maketitle
\section{Introduction}
In inelastic hadronic interactions with multiparticle production
only the fraction of collision energy is converted into the
production of secondaries. For the quantitative estimation of this
fraction one can use for a given interaction a characteristic,
originating from cosmic ray physics, inelasticity, which can be
defined as
\begin{equation}
k_{1}(s)=\frac{M}{\sqrt{s}},\label{inel1}%
\end{equation}
where $s$ -- is square of c.m. energy and $M$ is the mass of
intermediate system which decays into final produced particles.
The remaining part of incident energy is carried away by
participant's remnants -- so-called leading particles. From \
experimental point of view more suitable is another definition of
inelasticity
\begin{equation}
k_{2}(s)=\frac{1}{\sqrt{s}}\sum\limits_{i}\int
dy\mu_{i}\frac{dn_{i}}{dy}\cosh
y,\label{inel2}%
\end{equation}
where $\mu_{i}=\sqrt{p_{T_{i}}^{2}+m_{i}^{2}}$ is transverse mass
of a produced particle of type $i$\ and $dn_{i}/dy$ is its
measured rapidity distribution. \ Fluctuation of inelasticity from
event to event leads to the distribution $P(k)$ with mean
inelasticity $<k(s)>$. The energy dependence of inelasticity is a
problem of great interest both from theoretical and experimental
points of view. There is no consensus in the physical community on
the energy dependence of $<k(s)>$. The decrease of inelasticity
with energy is advocated by some authors \cite{fow, wlod, shab,
he} while others believe that inelasticity is an increasing
function of energy \cite{dias, gais, kop, wilk}. The question can
not be answered by collider experiments. At ISR energies (23
$\div$ 60 GeV) where leading particle spectrum could be measured
the inelasticity defined to be about 0.5. In the collider
experiments at higher energies (SPS and Tevatron) leading
particles are emitted in an extremely forward cone and could not
be measured due to the presence of the beam pipe. Obviously,
multiplicity distributions are connected with inelasticity
distributions, and so one can study features of multiplicity
distributions deriving the information on inelasticity or fraction
of the initial energy converted into the particle production. As
we know scaled\ multiplicity distributions exhibit KNO \cite{KNO}
scaling up to ISR energies which is violated for higher energy
data (SPS) where they can be described approximately by the
negative binomial distribution (NBD). And again at the most high
SPS energy, 900 GeV, there is an evident deviation of multiplicity
distributions from NBD. In this paper we demonstrate that there is
a connection between the energetic behaviour of the shapes of
scaled multiplicity distributions and inelasticity distributions
which, in turn, relates to the effect of the total cross section
growth. For this purpose we use geometrical considerations which
are justified by the following arguments. First, hadrons are
extended objects with the size about 1 fm and ,second, at high
energies de--Bloglie wave--length becomes small.

Our analysis is based on the model of hadron structure so-called
Strongly Correlated Quark Model (SCQM) \cite{Mus1} which is
described in Section 2. In Section 3 the model is applied for the
calculation of proton -- proton total cross sections and
simulation of inelastic events.

\section{Strongly Correlated Quark Model}

\ The ingredients of the model are the following. A single quark a
of definite colour embedded in vacuum starts to polarize its
surrounding that results in the formation of quark and gluon
condensate. At the same time it experiences the pressure of vacuum
because of zero point radiation field or vacuum fluctuations which
act the quark tending to destroy the ordering of the condensate.
Suppose that we place the corresponding antiquark in the vicinity
of the first one. Owing to their opposite signs colour
polarization fields of quark and antiquark interfere destructively
in the overlapped space regions eliminating each other at most in
the middle--point between the quarks. This effect leads to the
decreasing of condensates density in the same space region and
overbalancing of the vacuum pressure acting on quark and antiquark
from outer space regions. As a result the attractive force between
quark and antiquark emerges and quark and antiquark start to move
towards each other. The density of the remaining condensate around
quark (antiquark) is identified with the hadronic matter
distribution. At maximum displacement in $\overline{q}q-$ system
that corresponds to small overlapping of polarization fields the
hadronic matter distributions have maximum extent and magnitude.
The closer they to each other, the larger is the effect of mutual
destruction and the smaller hadronic matter distributions are
around quarks and the larger their kinetic energies. In that way
quark and antiquark start to oscillate around their middle--point.
For such interacting $\overline{q}q-$ pair located on $X$ axis
at the distance of $2x$ from each other the total Hamiltonian is%

\begin{equation}
H=\frac{m_{\overline{q}}}{(1-\beta^{2})^{1/2}}+\frac{m_{q}}{(1-\beta
^{2})^{1/2}}+V_{\overline{q}q}(2x),
\end{equation}
were $m_{\overline{q}}$, $m_{q}$ -- current masses of valence
antiquark and quark, $\beta=\beta(x)$ -- their velocity depending
on displacement $x$ and $V_{\overline{q}q}$ -- quark--antiquark
potential energy at separation $2x.$ It can be rewritten as
\begin{equation}
H=\left[  \frac{m_{\overline{q}}}{(1-\beta^{2})^{1/2}}+U(x)\right]
+\left[ \frac{m_{q}}{(1-\beta^{2})^{1/2}}+U(x)\right]
=H_{\overline{q}}+H_{q},
\end{equation}
were $U(x)=\frac{1}{2}V_{\overline{q}q}(2x)$ is potential energy
of quark or antiquark. Quark (antiquark) with the surrounding
cloud (condensate) of quark - antiquark pairs and gluons, or
hadronic matter distribution, forms the constituent quark. It is
natural to assume that the potential energy of quark (antiquark),
$U(x),$ corresponds to the mass $M_{Q}$ of the constituent quark:
\begin{equation}
2U(x)=C_{1}\int_{-\infty}^{\infty}dz^{\prime}\int_{-\infty}^{\infty}%
dy^{\prime}\int_{-\infty}^{\infty}dx^{\prime}\rho(x,{\mathbf{r}^{\prime}%
})\approx2M_{Q}(x)\label{poten}%
\end{equation}
where $C_{1}$ is a dimensional constant and hadronic matter
density distribution, $\rho(x,{\mathbf{r}^{\prime}}),$ is defined
as
\begin{equation}
\rho(x,{\mathbf{r}^{\prime}})=C_{2}\left|  \varphi(x,\mathbf{r}^{\prime
})\right|  =C_{2}\left|  \varphi_{Q}(x^{\prime}+x,y^{\prime},z^{\prime
})-\varphi_{\overline{Q}}(x^{\prime}-x,y^{\prime},z^{\prime})\right|  .
\end{equation}
Here $C_{2}$ is a constant, $\varphi_{Q}$ and
$\varphi_{\overline{Q}}$ are density profiles of the condensates
around quark and antiquark located at distance $2x$ from each
other. Here, we consider that the condensates around quark and
antiquark have opposite colour charges. They look like compressive
stress and tensile stress (around defects) in solids. The
generalization to a three-quark system in baryons is performed
according to $SU(3)_\mathrm{color}$ symmetry: in general, pair of
quarks have coupled representations
\begin{equation}
3\otimes3=6\oplus\overline{3}%
\end{equation}
in $SU(3)_\mathrm{color}$ and for quarks within the same baryon
only the $\overline{3}$ (antisymmetric) representation occurs. \
Hence, the antiquark can be replaced by two correspondingly
coloured quarks to get a colour singlet baryon and the destructive
interference takes place between colour fields of three valence
quarks (VQs). Putting aside the mass and charge differences and
spins of valence quarks we may say that inside the baryon three
quarks oscillate along the bisectors of the equilateral triangle.
Therefore, keeping in mind that quark and antiquark in mesons and
three quarks in baryons are strongly correlated, we can consider
each of them separately as undergoing oscillatory motion under the
potential (\ref{poten}) in 1+1 dimension. Hereinafter, we consider
VQ oscillating along $X-$ axis, and $Z-$ axis is a perpendicular
to the plane of oscillation $XY$. Density profiles of condensates
around VQs are taken in the gaussian form. It has been shown in
papers \cite{sch} that the wave packet solutions of time dependent
Schrodinger equation for a harmonic oscillator move in exactly the
same way as corresponding classical oscillators. These solutions
are called ``coherent states''. This relationship justifies our
semiclassical treatment of quark dynamics.

\begin{figure}
\includegraphics[width=5in]{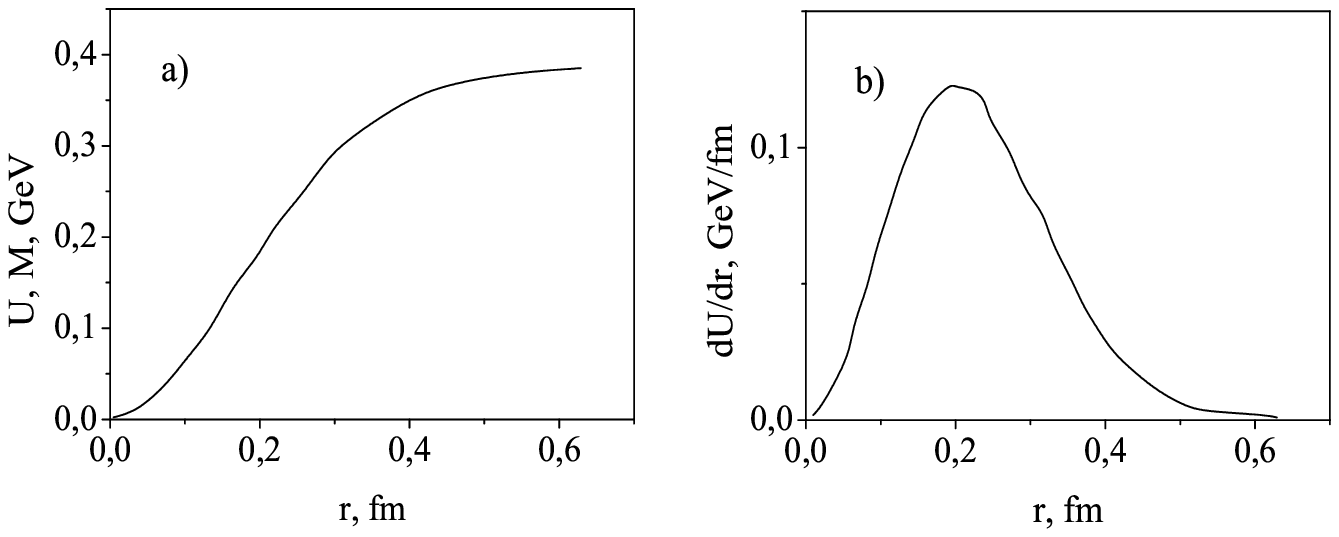}
\caption{ a) Potential energy of valence quark and mass of constituent
quark; b) ''Confinement'' force.}
\label{potfig}%
\end{figure}

\begin{figure}
\includegraphics[width=6in]{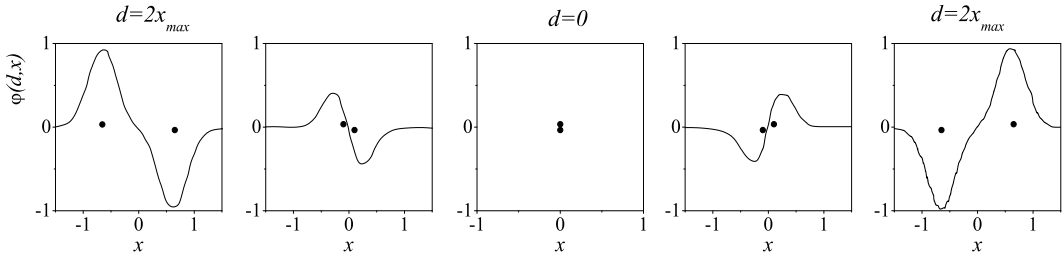}
\caption{Evolution of color charge density profile, $\varphi,$ in
quark--antiquark system during half--period of oscillations; $d=2x$ --
distance in \textit{fermi} between quark and antiquark depicted as dots.}%
\label{scqm}%
\end{figure}

We specify the mass of constituent quark at maximum displacement as
\[
M_{Q(\overline{Q})}(x_{\max})=\frac{1}{3}\left(  \frac{m_{\Delta}+m_{N}}%
{2}\right)  \approx360 \ \mathrm{MeV},
\]
where $m_{\Delta}$ and $m_{N}$ are masses delta--isobar and
nucleon respectively. The parameters of the model, namely, maximum
displacement, $x_{\max},$ and parameters of the gaussian function,
$\sigma_{x,y,z},$ for hadronic matter distribution around VQ are
chosen inside the following corridors:
\begin{equation}
x_{\max}=0.64\div0.66 \ \mathrm{fm},\ \sigma_{x,y}=0.24\div0.28 \
\mathrm{fm},\ \sigma
_{z}=0.12\div0.20 \ \mathrm{fm}.\label{params}%
\end{equation}
They are estimated by comparison of calculated and experimental
values of inelastic cross sections, $\sigma_\mathrm{in}(s),$ and
the inelastic overlap function $G_\mathrm{in}(s,b)$ for $pp-$ and
$\overline{p}p-$ collisions (see the next section). The current
mass of the valence quark is taken to be $5 \ \mathrm{MeV}$. The
behaviour of potential (\ref{poten}) evidently demonstrates the
relationship between constituent and current quark states inside a
hadron (Fig. \ref{potfig}). At maximum displacement quark is
nonrelativistic, constituent one (VQ surrounded by the
condensate), since the influence of polarization fields of other
quarks becomes minimal and the VQ possesses the maximal potential
energy corresponding to the mass of the constituent quark. At the
origin of oscillation, $x=0,$ antiquark and quark in mesons and
three quarks in baryons, being close to each other, have maximum
kinetic energy and correspondingly minimum potential energy and
mass: they are relativistic, current quarks (bare VQs). This
configuration corresponds to so-called ``asymptotic freedom'' . In
the intermediate region there is  increasing (decreasing) of the
constituent quark mass by dressing (undressing) of VQs due to
decreasing (increasing) of the destructive interference effect.
The evolution of colour charge density profiles of quark --
antiquark pair during the half-period of oscillation is shown in \
Fig. \ref{scqm}. Here, we suppose that the quark colour charge is
positive and the antiquark color charge is negative.

The proposed dynamical picture meets local gauge invariance
principle. Indeed, destructive interference of color fields of
quark and antiquark in mesons and three quarks in baryons
depending on their displacements can be treated as a phase
rotation of wave function of a single VQ in colour space
$\psi_{c}$ on angle $\theta$ depending on the displacement $x$ of
the VQ in the coordinate space
\begin{equation}
\psi_{c}(x)\rightarrow e^{ig\theta(x)}\psi_{c}(x).\label{gauge1}%
\end{equation}
The color phase rotation,in turn, leads  to VQ dressing
(undressing) by quark and gluon condensate that corresponds to the
transformation of a gauge field
\begin{equation}
A_{\mu}(x)\rightarrow A_{\mu}(x)+\partial_{\mu}\theta(x).\label{gauge2}%
\end{equation}
Here, we drop colour indices of $A_{\mu}(x)$ and consider each
quark of specific colour separately as changing its effective
colour charge, $g\theta(x),$ in color fields of other quarks
(antiquark) due to the destructive interference. Thus, gauge
transformations (\ref{gauge1}), (\ref{gauge2}) map internal
(isotopic) space of a colored quark onto the coordinate space. On
the other hand, this dynamical picture of VQ dressing (undressing)
corresponds to the chiral symmetry breaking (restoration). Due to
this mechanism of VQs oscillations the nucleon runs over the
states corresponding to the certain terms of the infinite series
of Fock space
\begin{equation}
\mid B\rangle=c_{1}\mid q_{1}q_{2}q_{3}\rangle+c_{2}\mid q_{1}q_{2}%
q_{3}\overline{q}q\rangle+c_{3}\mid q_{1}q_{2}q_{3}g\rangle...
\end{equation}
The proposed model has some important consequences. Inside hadrons
the valence quarks and accompanying them gluons and
quark-antiquark pairs, as well, are strongly correlated. Nucleons
are nonspherical object: they are flattened along the axis
perpendicular to the plane of quarks oscillations.

\begin{figure}
\includegraphics[width=5in]{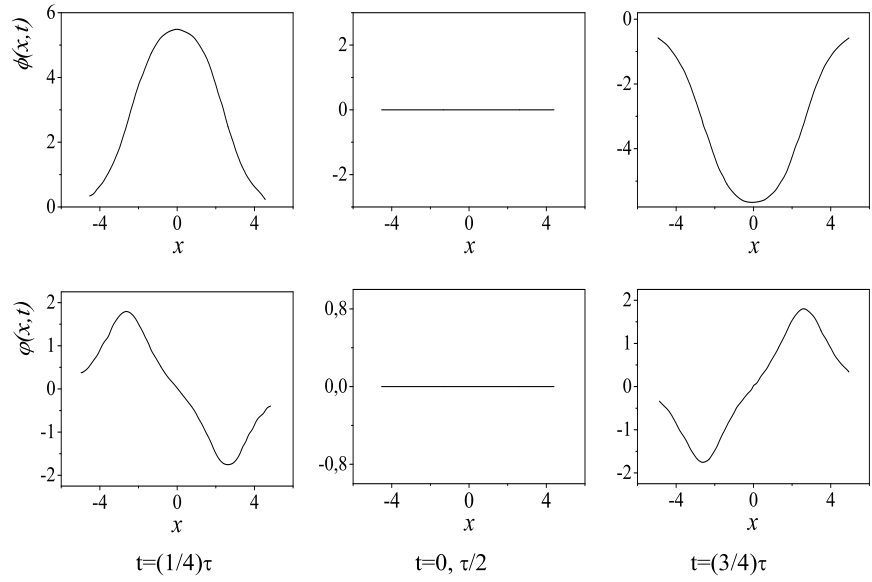}
\caption{Evolution of breather, $\protect\phi ,$ and its energy density
profile, $\protect\varphi ,$ during a half--period of oscillation. Scales
are abitrary.}\label{breather}%
\end{figure}

From the form of the quark potential (Fig. \ref{potfig})\ one can
conclude that dynamics of VQ corresponds to nonlinear oscillator
and VQ with its surrounding can be treated as a nonlinear wave
packet. Moreover, our quark -- antiquark system turned out to be
identical to so-called  ``breather''  solution of (nonlinear)
sine--Gordon (SG) equation\cite{Mus2}. SG equation in (1+1)
dimension in the reduced form for scalar function $\phi(x,t)$ is
given by
\begin{equation}
\Box\phi(x,t)+\sin\phi(x,t)=0,
\end{equation}
where $x,$ $t$-dimensionless. Breather is a periodic solution
representing a bound state of a soliton-antisoliton pair which
oscillates around its center of mass:
\begin{equation}
\phi_{br}(x,t)=4\tan^{-1}\left[  \frac{\sinh\left[
ut/\sqrt{(1-u^{2}}\right] }{u\cosh\left[  x/\sqrt{(1-u^{2}}\right]
}\right]  ,
\end{equation}
where $u$ is 4-velocity. During the oscillations of the
soliton-antisoliton pair their density profile
\begin{equation}
\varphi_{br}(x,t)=\frac{d\phi_{br}(x,t)}{dx}%
\end{equation}
evolves like our quark--antiquark system, i.e. at the maximal
displacement the soliton and antisoliton are emphasized maximal
and at the minimum displacement they  ``annihilate''  (Fig.
\ref{breather}). This similarity is not surprising because our
quark--antiquark system was formulated in close analogy with the
model of dislocation-antidislocation which in the continuous limit
is described by the breather solution of SG equation \cite{raja}.
It can be shown that soliton, antisoliton and breather obey
relativistic kinematics, i.e. their energies, momenta and shapes
are transformed according to Lorentz transformations. Since the
above consideration of quarks as solitons is purely classical the
important problem is to construct \ quantum states around them.
Although the soliton solution of SG equation looks like an
extended (quantum) particle the relation of classical solitons to
quantum particles is not so trivial. The technique of quantization
of the classical solitons with the usage of various methods has
been developed by many authors. The most known of them is
semiclassical method of quantization (WKB) which allows one to
relate classical periodic orbits (breather solution of SG) with
the quantum energy levels \cite{DHN}.

Hereinafter, we adhere to our semiclassical model (SCQM) applying
it for analysis of cross sections and the multiparticle production
in hadron--hadron collisions.

\section{Hadron - Hadron Collisions}

\begin{figure}
\includegraphics[width=4.5in]{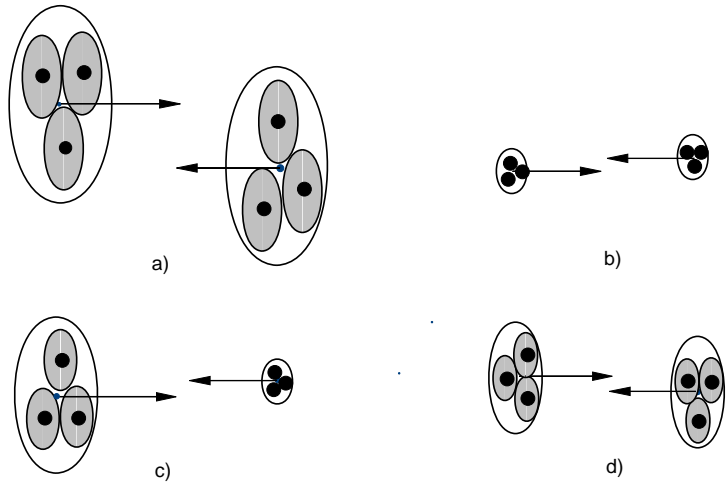}
\caption{Different quark configurations realized inside colliding nucleons
at
the instant of collision.}%
\label{q-config}%
\end{figure}

Different configurations of the quark content inside a hadron
realized at the instant of the collision result in different types
of reactions. The probability of finding any quark configuration
inside the hadron is defined by the probability of VQ's
displacement in a proper frame of the hadron:
\begin{equation}
P(x)dx=\frac{Adx}{\sqrt{1-m_{q}^{2}/\left[  E_{q}-U(x)\right]  ^{2}}%
},\label{cnfprob}%
\end{equation}
where $E_{q}$ is a total energy of the valence quark (antiquark)
and a constant\ $A$ can be derived from the normalization
condition
\begin{equation}
\int\limits_{-\infty}^{\infty}P(x)dx=1.
\end{equation}
Configurations with nonrelativistic constituent quarks ( $x\simeq
x_\mathrm{max}$) in both colliding hadrons lead to soft
interactions with the nondiffractive multiparticle production in
central and fragmentation regions (Fig. \ref{q-config}a). The hard
scattering with the jet production and the large angle elastic
scattering take place when configurations with current VQs
($x\simeq0$) in both colliding hadrons are realized (Fig.
\ref{q-config}b). The near current quark configuration inside one
of the hadrons and constituent quark configuration inside the
second one result in single diffraction scattering (Fig.
\ref{q-config}c). And at last intermediate configurations inside
one or both hadrons are responsible for semihard and double
diffractive scattering (Fig. \ref{q-config}d). The same
geometrical consideration can be applied to deep inelastic
scattering processes if one assumes that a real or virtual photon
converts into the vector meson according to the vector dominance
model.

First, we apply our model for the calculation of proton-proton and
antiproton-proton cross sections at high energies and demonstrate
that the growth of the cross section with energy is caused by
predominantly increasing contribution of peripheral interactions
that, in turn, leads to the decreasing of inelasticity of
collisions. Then we will show that the energetic behaviour of
inelasticity distributions\ governs the energetic behaviour of
scaled multiplicity distributions.

\subsection{Cross Sections}

To calculate cross sections we used an impact parameter
representation, namely Inelastic Overlap Function (IOF), which can
be specified via the unitarity equation
\begin{equation}
2Imf(s,b)=|f(s,b)|^{2}+G_\mathrm{in}(s,b),
\end{equation}
where $f(s,b)-$elastic scattering amplitude and
$G_\mathrm{in}(s,b)$ is IOF. IOF is connected with inelastic
differential cross sections in impact parameter space:
\begin{equation}
\frac{1}{\pi}(d\sigma_\mathrm{in}/db^{2})=G_\mathrm{in}(s,b).\label{Gin}%
\end{equation}
Then the inelastic, elastic and total cross sections can be
expressed via IOF as
\begin{equation}
\sigma_\mathrm{in}(s)=\int
G_\mathrm{in}(s,\mathbf{b})d^{2}\mathbf{b},
\end{equation}%
\begin{equation}
\sigma_{el}(s)=\int\left[  1-\sqrt{1-G_\mathrm{in}(s,\mathbf{b})}\right]  ^{2}%
d^{2}\mathbf{b},
\end{equation}%
\begin{equation}
\sigma_\mathrm{tot}(s)=2\int\left[
1-\sqrt{1-G_\mathrm{in}(s,\mathbf{b})}\right]
d^{2}\mathbf{b}.\label{sigtot}%
\end{equation}
Since IOF relates to the probability of inelastic interaction at a
given impact parameter, we carried out Monte Carlo simulation of
inelastic nucleon--nucleon interactions. The Inelastic interaction
takes place at the definite impact parameter $b$ if the produced
mass meets the following requirement:
\begin{equation}
M_{CF}^{2}=4M_{P}\gamma_{P}M_{T}\gamma_{T}\int\rho_{P}(\mathbf{r})\rho
_{T}(\mathbf{r}-\mathbf{b})~d^{3}\mathbf{r}\geq(M_{CF}^{2})_{\min
},\label{heis}%
\end{equation}
where $M_{CF}$ is the mass of the central ``fireball'' produced at
the ovelapped region, $\rho_{P}$ and $\rho_{T}$\ are hadronic
matter density distributions in projectile and target hadrons,
$M_{P}$, $M_{T}$ -- masses, $\gamma_{P}$, $\gamma_{T}$ --
gamma--factors of the colliding hadrons and $(M_{CF}^{2})_{\min}$
-- the minimal mass of a fireball that results in an inelastic
event. This expression is the modification of assumption of
Heisenberg \cite{heisen} on interaction of extended particles: we
replaced in his original formula the pion mass squared (on the
right hand side) by $(M_{CF}^{2})_{\min}$. In our previous papers
this
quantity corresponded to the transverse mass of the pion: $m_{\pi_{\perp}}%
^{2}=p_{\perp}^{2}+m_{\pi}^{2}.$ Taking into account the energy
dependence of the average momentum of produced particles and the
increasing yield of minijets (as treated in what follows) we
parametrize the minimal fireball mass as
\begin{equation}
(M_{CF})_{\min}=0.3+0.03s^{1/4}.\label{minms}%
\end{equation}

\begin{figure}
\includegraphics[width=4.7in]{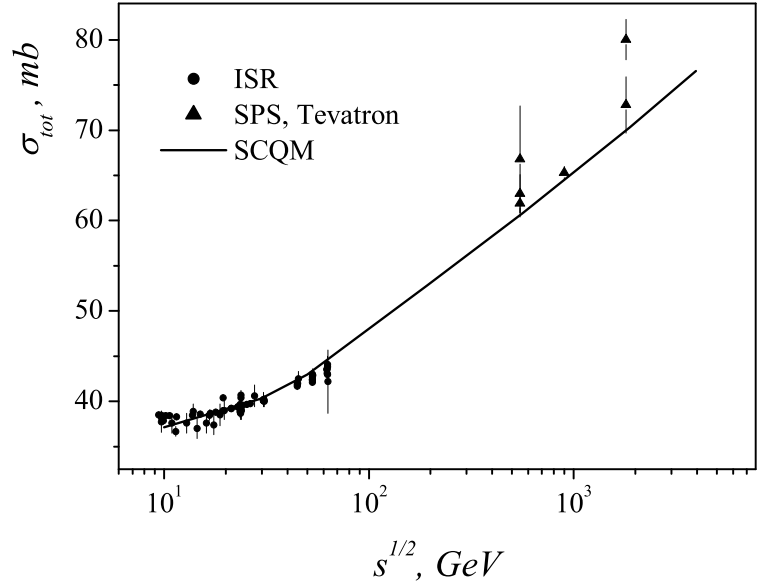}
\caption{Total cross section for $pp$ and $\overline{p}p$. Data are taken
from
electronic data base HEPDATA \cite{hep}.}%
\label{totcs}%
\end{figure}

\begin{figure}
\includegraphics[width=5in]{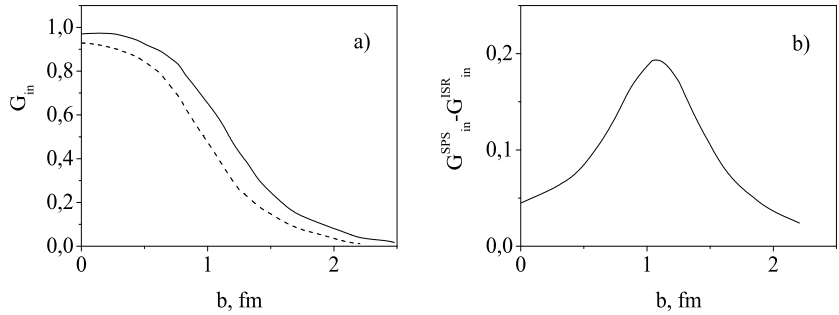}
\caption{Plot of overlap functions at ISR and SPS energies, (a), and the
difference of overlap functions $G_{in}^{SPS}-G_{in}^{ISR},$ (b), as a
function of impact parameter $b.$ }%
\label{GinDif}%
\end{figure}

Specifying the quark configurations in each colliding hadrons
according to (\ref{cnfprob}) we calculated $G_\mathrm{in}(s,b)$
for particular values of the impact parameter $b$ and then
according to (\ref{sigtot}) -- total cross sections,
$\sigma_\mathrm{tot}$. Fig. \ref{totcs} shows the results of the
calculation for total cross sections for proton -- proton and
antiproton -- proton collisions in a wide range of collision
energies. One can see that the model with fixed parameters
characterizing the geometrical size of hadrons describes the
energetic behaviour of $\sigma_{tot}$ rather well$.$ The growth of
the total cross section with energy coming from the growth of the
inelastic cross section is due to the continuous tails of
condensates (hadronic matter distributions) around VQs not
compensated by the destructive interference effect inside each
interacting particle. With rising collision energy the overlap of
more peripheral parts of these tails make it possible to meet
requirement (\ref{heis}) and consequently results in the
increasing effective size of the hadronic matter distribution
inside nucleons and correspondingly the increasing radius of
interactions. It can be seen from the comparison of IOFs for ISR
and SPS energies which is given in Fig. \ref{GinDif}. According to
Eq. (\ref{Gin}) the difference
$G_\mathrm{in}^{SPS}-G_\mathrm{in}^{ISR}$ exhibits the
predominantly peripheral increase of the inelastic cross section
(and thus of the total cross section, since $\sigma_{el}/$
$\sigma_{tot}$ is only about 20\%) which is centered around 1 fm
(Fig. \ref{GinDif}b). As noted by the authors of the paper
\cite{henzi} at high energies colliding nucleons become blacker,
edger and larger (``BEL--effect''). The model gives the linear
logarithmic energy dependence for total cross sections. At
energies $\sqrt{s}<30$ $GeV$ calculated cross sections were
corrected on contributions of Regge poles exchange by using
Donnachie and Landshoff parametrization\cite{donlans}. An
oscillatory motion of VQs appearing as an interplay between
constituent and bare (current) quark configurations results in
fluctuations of the hadronic matter distribution inside colliding
nucleons. The manifestation of these fluctuations is a variety of
scattering processes, hard and soft, in particular, the process of
single diffraction (SD). SD--events correspond to the constituent
quark configuration inside one colliding hadron and (semi)bare
quark configuration inside the other one. Our unified geometrical
explanation of diffractive, nondiffractive and DIS processes could
give an answer to the long standing question: what is pomeron?
Historically the concept of ``Pomeron'' originating from simple
Regge pole with the intercept $a_{0}=1$ transformed to a rather
complicated object with relatively arbitrary features and smooth
meaning. To produce the rising cross sections it must have the
intercept such that $a_{0}=1+\varepsilon$. The fact that the
parameter $\varepsilon$ is universal, independent of particles
being scattered in hadronic and DIS interactions, could say us
that the nature of the cross section growth is the same for all
processes. Our interpretation of pomeron is a geometrical one.
Both diffractive and nondiffractive particle productions emerge
from the disturbance (excitation) of overlapped continuous vacuum
polarization fields (gluon and $\overline{q}q$ condensate) around
valence quarks of colliding hadrons followed by fragmentation
process. The type of the interaction depends on quark
configurations inside a colliding hadron occurring at the instant
of the interaction and the value of the impact parameter. So, what
we used to call ``Pomeron''  in $t-$ channel is solely continuum
states in $s-$ channel and we claim that Pomeron is unique in
elastic, inelastic (diffractive and nondiffractive) and DIS.

\subsection{Multiparticle Production in Hadronic Collisions}

According to our model the configurations with nonrelativistic
constituent quarks ( $x\simeq x_{max}$) inside the both colliding
hadrons lead to soft interactions with multiparticle production in
central and fragmentation regions. The additional restriction by
small impact parameters selects central collisions when hadronic
matter distributions of colliding hadrons (quarks) overlap
totally. In this case kinetic energies of colliding hadrons
dissipate totally converting into the production of secondary
particles that corresponds to collision inelasticity close to 1
and very high multiplicity in comparison with the mean one. We
will consider the soft interactions and nondiffractive
multiparticle production, in particular. According to the KNO
hypothesis the scaled multiplicity distributions, $\left\langle
n\right\rangle P_{n}(s)$, depend on the ratio of the number of
particles to the average multiplicity $z=n/\left\langle
n\right\rangle $ and they are energy independent. From our
geometrical point of view such behaviour could be explained as a
superposition of relatively narrow distributions corresponding to
the particular impact parameters of the collisions. Indeed, the
multiplicity distribution can be defined as
\begin{equation}
P_{n}(s)=\int_{0}^{1}P(n\mid k)\text{ }P[k(s)]dk,\label{conmult}%
\end{equation}
where $P[k(s)]$ is inelasticity distribution and $P(n\mid k)$ is
the probability of the production of $n$ particles at the given
inelasticity, $k.$\ So, if the conditional probability $P(n\mid
k)$ is sufficiently narrow then the shape of the distribution
$P_{n}$ is defined by the shape of the inelasticity distribution
$P[k(s)].$ The inelasticity distributions are strictly connected
with the impact parameter distributions. KNO scaling holds (at
least, approximately) if the impact parameter distributions and
consequently inelasticity distributions are energy independent. As
shown in the previous subsection the growth of inelastic and total
cross sections with energy in hadronic collisions is due to the
increasing of effective sizes of interacting hadrons. To make a
quantitative analysis of energetic dependence of multiplicity
distributions we performed, in the frame of SCQM, Monte Carlo
simulation of inelastic proton -- proton interactions selecting
nondiffractive events. The process of simulation includes the
following steps.

\begin{figure}
\includegraphics[width=3.in]{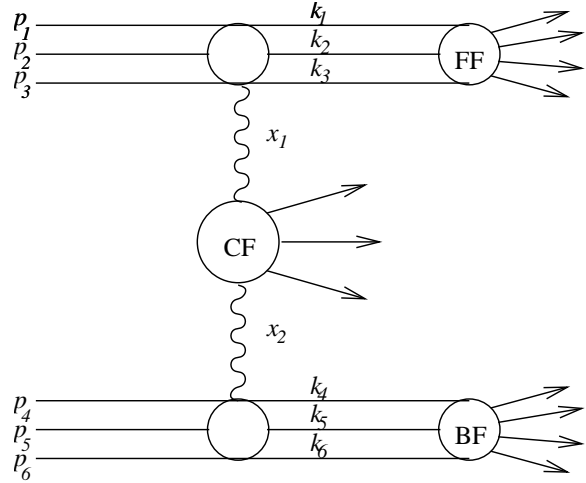}
\caption{Fireball picture of multiparticle production in proton -- proton
collision. $p_{i}$ and $k_{i}$ are momenta of constituent quarks before
and after collision, respectively; $x_{1}$ and $x_{2}$ -- fractions of
protons momenta
forming the central fireball. }%
\label{ppdiag}%
\end{figure}

\begin{figure}
\includegraphics[width=4.in]{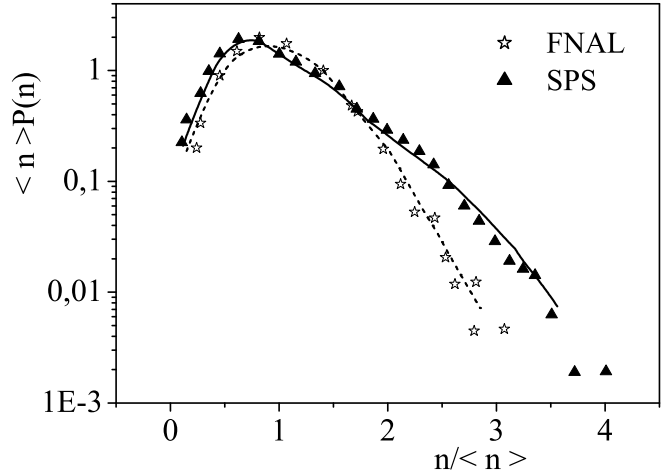}
\caption{ The multiplicity distributions of charged particles in $pp-$ and
$\overline{p}p-$ collisions. Data are taken from
electronic data base HEPDATA \cite{hep}. }%
\label{multdis}%
\end{figure}

\begin{figure}
\includegraphics[width=4.in]{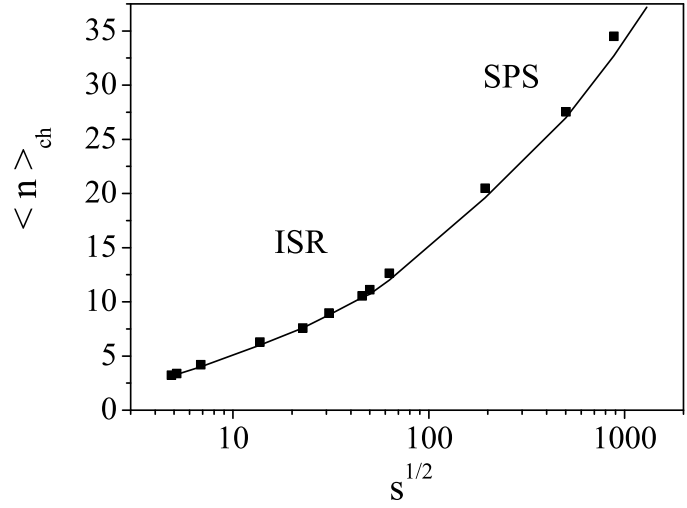}
 \caption{The energetic dependence of the average
multiplicity of charged particles in $pp-$ and $\overline{p}p-$
collisions. Data are taken from
electronic data base HEPDATA \cite{hep}. }%
\label{amult}%
\end{figure}

\begin{figure}
\includegraphics[width=4.in]{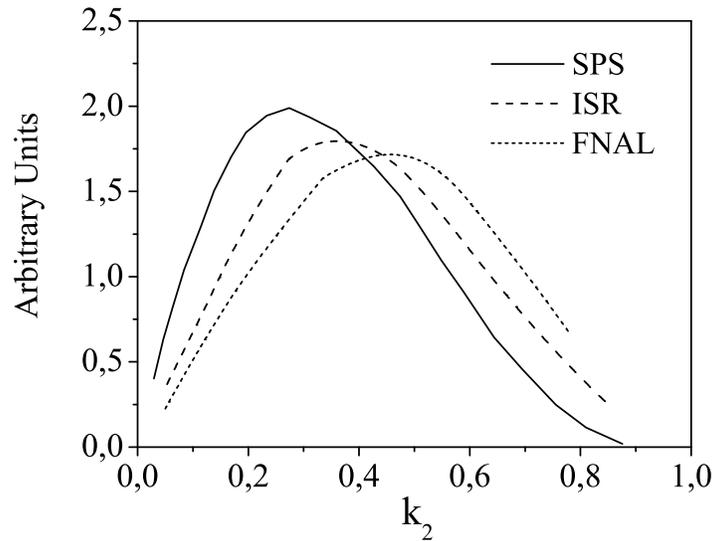}
\caption{ Inelasticity distributions for $pp-$ and $\overline{p}p-$
collisions calculated according to Eq. (\ref{inel2}).} \label{ineldis}
\end{figure}

\begin{figure}
\includegraphics[width=4.in]{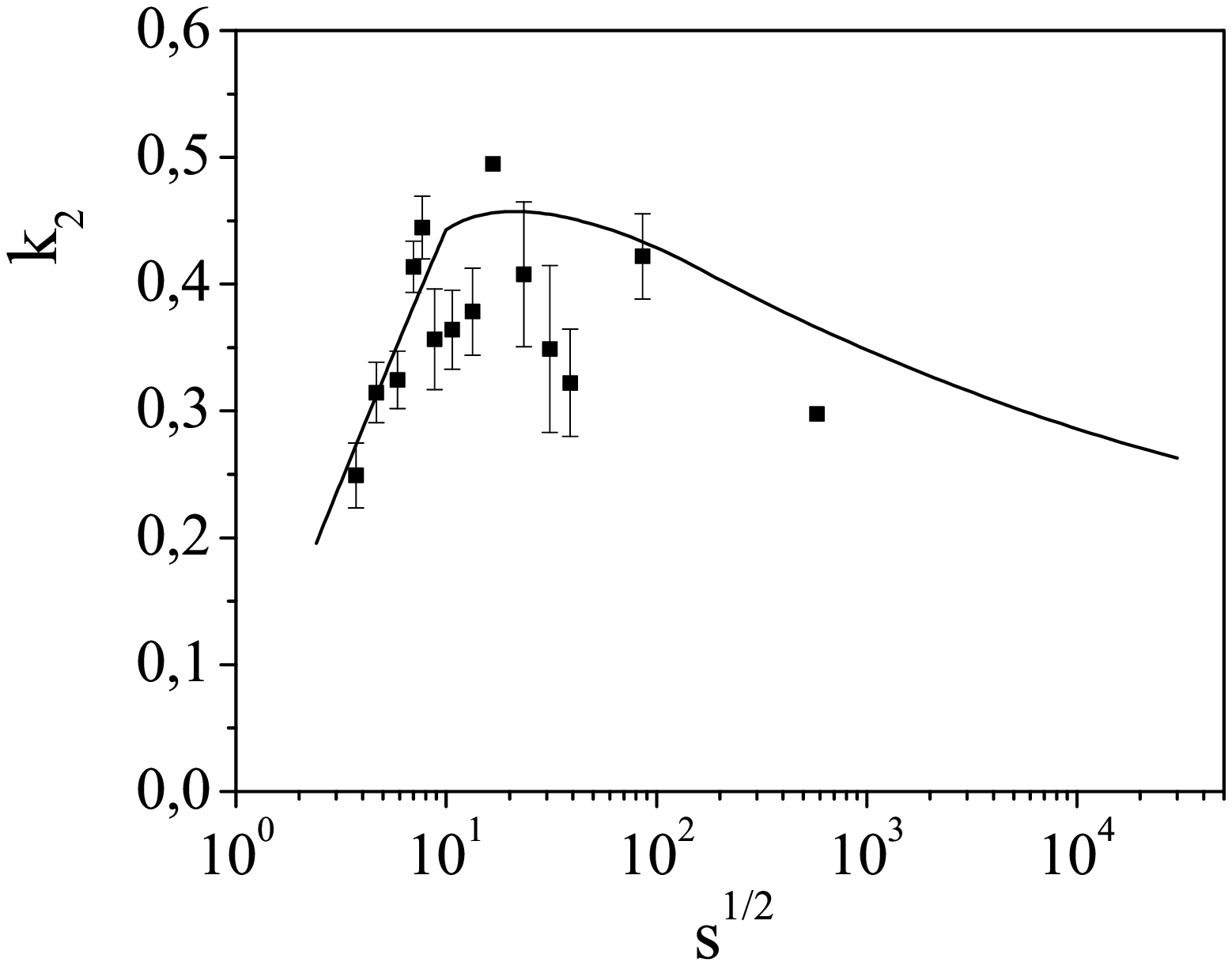}
\caption{The average inelasiticity plotted as a function of the energy for
$pp-$ and $\overline{p}p-$ collisions. Boxes are compilation of data given
in
paper \cite{barash}.}%
\label{ainel}%
\end{figure}

\begin{enumerate}
\item  Applying Heisenberg prescription (\ref{heis}) we define the mass of
the central ``fireball'' (CF) (Fig. \ref{ppdiag}) produced in the
proton--proton collision at a particular impact parameter. Quark
configurations inside each proton at the instant of the collision
is specified randomly according to the probability (\ref{cnfprob})
that allows one to fix energies and momenta of quarks inside both
protons. Since the mass of CF is formed by the overlap of hadronic
densities of individual constituent quarks (CQ) of colliding
protons we know the energies and momenta of quarks in both
remnants which we call, by convention, forward and backward
``fireballs''  (FF and BF). The notion ``fireball''  is applied by
convention only because all fireballs, CF, FF and BF, can decay
string -- like manner and there is no sharp boundary between
secondaries emitted from fireballs in the rapidity space for
nondiffractive
events. Then we calculate the effective masses of FF\ and BF (Fig. \ref{ppdiag}):%

\begin{eqnarray}
M_{FF}  &=&\sqrt{\left(  \sum\limits_{i=1}^{3}E_{i}^{^{\prime}}\right)
^{2}-\left(  \sum\limits_{i=1}^{3}\mathbf{k}_{i}\right)  ^{2},}\\ M_{BF}
&=&\sqrt{\left(  \sum\limits_{i=4}^{6}E_{i}^{^{\prime}}\right) ^{2}-\left(
\sum\limits_{i=4}^{6}\mathbf{k}_{i}\right)  ^{2}},
\end{eqnarray}
where $E_{i}^{^{\prime}}$ and $\mathbf{k}_{i}$ are energies and
momenta of constituent quarks after the collision.

\item  We assume that\ each fireball breaks up, in general, into clusters.
Here, the bremsstrahlung analogy is used, namely, at the instant
of the collision a proton (electron) loses the energy dumping
fraction of its hadronic (electromagnetic) field by means of the
emission of clusters (photons). To simulate the masses of the
clusters we apply the result of the paper \cite{chou} for a
cluster mass spectrum
\begin{equation}
P(m_{cl})=(m_{cl}/m_{0})\exp(-m_{cl}/m_{0}),
\end{equation}
following from the statistical nature of the cluster emission. Our
next assumption is that masses of clusters increase with the
collision energy. This is dictated by the necessity to take into
account such peculiarities of multiparticle production as the
growth of rapidity distribution plateau, the increasing of
transverse momenta of secondaries and the increasing yield of
minijets. We parametrize the energy dependence of the average mass
of clusters as
\begin{equation}
\left\langle m_{cl}\right\rangle =0.3+0.09s^{1/4}.
\end{equation}
Notice that we have chosen the same energetic dependence for the
minimal fireball mass in Heisenberg prescription (\ref{minms})
except for the value of the slope parameter.

\item  Given the positions of the centers of masses of each fireball in rapidity
space and kinematically allowed rapidity (sub)spaces for the
breaking up of each fireball into clusters we simulate the momenta
for each generated cluster in the proper frame of the
corresponding fireball. Bremsstrahlung mechanism of the fireball
fragmentation corresponds to statistically independent emission of
clusters with the limited transverse momenta. Therefore, we apply
the cylindrical phase space model according to which the rapidity
of \textit{i}-th cluster is defined as
\begin{equation}
y_{i}=\xi Y_{i},
\end{equation}
where $\xi$ is random number uniformly distributed in the interval
[0,1]. The allowed rapidity interval, $Y_{i},$ is given by
\[
Y_{i}=\ln(M_{F}^{2}/(\mu_{cl})_{i}^{2},
\]
where $(\mu_{cl})_{i}^{2}=$ \ $(m_{cl})_{i}^{2}+p_{\bot i}^{2},$
transverse mass of the cluster $i.$ Moreover, the rapidity
interval for fragmentation of the central fireball, $Y_{i}^{CF},$
is restricted by the requirement
\begin{equation}
Y_{i}^{CF}\leq Y^{FF}-Y^{BF},
\end{equation}
where $Y^{FF}$ and \ $Y^{BF}$ are rapidities of forward and
backward fireball respectively.\ The transverse momenta of the
clusters are generated according to the distribution
\begin{equation}
f(p_{\bot}^{2})\propto\exp(-bp_{\bot}^{2}).
\end{equation}
The energy of the remnant baryon in the proper frame of FF (BF) is
defined by the summary energy of two quarks closest to each other
in rapidity space, i.e. kinematic characteristics of the baryon
are connected to those of ``diquark''.

\item  Since our clusters are identified with minijets they should decay in
a jet-like manner. One could assume that these minijets are formed
by the fragmentation of the excited sea quark--antiquark pair.
Hence, we can approximate the spectrum of cluster decay using data
on the electron--positron annihilation provided that the cluster
mass is identified with the center of mass energy of an electron
and positron: $m_{cl}=\sqrt{s}_{e^{+}e^{-}}$. One can apply for
this purpose any of appropriate Monte -- Carlo generators. The
axis of the decaying jet is generated to be directed
isotropically. And, at last, the model meets energy -- momentum
conservation requirements for all products of a reaction.

\qquad Calculated in such a manner the multiplicity distributions
in the KNO form and the energetic dependence of the mean
multiplicity for charged particles are shown in Figures
\ref{multdis} and \ref{amult}. Given all characteristics of
produced particles in the event we can calculate inelasticity
$k_{2}$ (Eq. \ref{inel2}). Its distributions for $pp-$
interactions at different collision energies are shown in Fig.
\ref{ineldis}. The inelasticity distribution evolves with energy
in such a way that its maximum position shifts to the lower values
of inelasticity at higher collision energies. It means that the
higher is the collision energy the lower is the average
inelasticity (Fig. \ref{ainel}). Analyzing the multiplicity
distribution at different energies one can see that its maximum
position shifts to the lower values of scaled multiplicities and
the contribution of high multiplicities increases while the
collision energy increases. According to Eq. (\ref{conmult}) the
multiplicity distribution can be expressed via the conditional
multiplicity distribution at particular inelasticity and the
inelasticity distribution. The conditional multiplicity
distribution at the particular inelasticity, in turn, is built
from multiplicities of clusters emitted from forward, backward and
central fireballs and multiplicities going from the clusters'
fragmentation. If the inelasticity distribution and (average) mass
of the clusters would not depend on the collision energy then the
scaled multiplicity distributions do not depend on energy either
and KNO - scaling takes place. The shift of the position of the
maximum of the inelasticity distribution with the energy growth
shifts the position of the scaled multiplicity distribution. On
the other hand, masses of clusters growing with collision energy
lead to the narrowing of available rapidity space and,
consequently, to the violation of Feynman scaling. This effect
most obviously exhibiting at the inelasticities close to 1 cause
the lift of a multiplicity distribution tail at high
multiplicities. To summarize we claim that the mean inelasticity
decreases with energy, and the violation of KNO - scaling is a
consequence of the growth of inelastic and total cross
sections and of masses of emitted clusters with energy. \\
This research was partly supported by the Russian Foundation of
Basics Research, grant 01--07--90144.
\end{enumerate}

\newpage
%
\newpage
%
\newpage
%
\newpage
%
\newpage
%
\newpage
%
\newpage
%

\newpage
%
\newpage
%
\newpage
%
\newpage
%


\newpage
\begin{thebibliography}{99}

\bibitem{fow}  G. Fowler, R. M. Weiner, G. Wilk, Phys. Rev. Lett., \textbf{55%
} (1985) 177.

\bibitem{wlod}  Z. Wlodarczyk, J. Phys. G: Nucl. Part. Phys., \textbf{19}
(1993) L128.

\bibitem{shab}  Yu. A. Shabelski \textit{et al}., J. Phys. G: Nucl. Part.
Phys., \textbf{18} (1992) 1281.

\bibitem{he}  Y. D. He, J. Phys. G: Nucl. Part. Phys., \textbf{19} (1993)
1953.

\bibitem{dias}  J. Dias de Deus, Phys. Rev. D 32 (1985) 334.

\bibitem{gais}  Gaisser T. K. and Stanev T., Phys. Lett., B219 (1989) 375.

\bibitem{kop}  Kopeliovich B. Z., Nikolaev N. N. and Potashikova I. K.,
Phys. Rev. D39 (1989) 769.

\bibitem{wilk}  F. S. Duraes, F. O. Navarra and G. Wilk, Phys. Rev., D47
(1993) 3049.

\bibitem{KNO}  Z. Koba, N. B. Nielsen and P. Olesen, Nucl. Phys., \textbf{B40%
} (1972) 317

\bibitem{Mus1}  G. Musulmanbekov, \textit{Proc. VIIIth Blois Workshop, }Ed.
V. A. Petrov, World Scientific, 2000, p. 341--350, and references therein.

\bibitem{sch}  E. Schrodinger, Naturwissenschaften \textbf{14} (1926) 664.

\bibitem{Mus2}  G. Musulmanbekov in \textit{Frontiers of Fundamental Physics
4}, Ed. B. G. Sidharth, Klewver Acad. Press, 2001, p. 109--120.

\bibitem{raja}  R. Rajaraman, Phys. Rep. \textbf{21C} (1975) 229.

\bibitem{DHN}  R. Dashen, B.Hasslacher and a. Neveu, Phys. Rev. \textbf{D10 }%
(1974) 4114.

\bibitem{heisen}  W. Heisenberg, Zeit. Phys., Bd. 133 (1952) 65.

\bibitem{hep}  http://durpdg.durham.ac.uk/HEPDATA.

\bibitem{henzi}  R. Henzi and P. Valin, Phys. Lett\textit{. }\textbf{132B }%
(1983) 443; R. Henzi,\textit{\ }Proc. of the 4th Topical Workshop on $%
\overline{p}p$\ Collider Physics, Bern, 1984.

\bibitem{donlans}  A.Donnachie and P.V. Landshoff, CERN--TH 6635/92.

\bibitem{chou}  Chou Kuang--chao, Liu Lian--son, Meng Ta--chung, Phys. Rev.,
\textbf{D28} (1983) 1080.

\bibitem{barash}  V. S. Barashenkov, N. B. Slavin, Acta Phys. Pol., \textbf{%
B12} (1981) 563.
\end{thebibliography}
\end{document}